\documentclass[preprint,superscriptaddress,showpacs]{revtex4}     
\pdfoutput=1

\usepackage[utf8]{inputenc}

\usepackage{graphicx}         
\usepackage{amssymb}          

\usepackage{soul}      

\usepackage[usenames,dvipsnames]{color}

\usepackage{siunitx}
\newcommand{\angs}{\angstrom}
\newcommand{\Ang}{\si{\angstrom}}

\usepackage{subscript}

\newcommand{\sups}[1]{\textsuperscript{#1}}
\newcommand{\subs}[1]{\textsubscript{#1}}

\newcommand{\+}{\sups{+}}
\newcommand{\n}{\subs{n} }

\begin{document}

\title{Path Integral Monte Carlo study confirms a highly ordered snowball in \4He nanodroplets doped with an Ar\+ ion
}

\author{Filippo Tramonto}
\email[Email: ]{filippo.tramonto@unimi.it}
\affiliation{Dipartimento di Fisica, Università degli Studi di Milano, via Celoria 16, 20133 Milano, Italy}
\author{Paolo Salvestrini}
\affiliation{CNR-IFN - Sezione di Milano, P.zza L. da Vinci, 32, 20133 Milano, Italy}
\author{Marco Nava}
\affiliation{Computational Science, Department of Chemistry and Applied Biosciences, ETH Zurich, Campus, Via Giuseppe Buffi 13, CH-6900 Lugano, Switzerland}
\author{Davide E. Galli} 
\affiliation{Dipartimento di Fisica, Università degli Studi di Milano, via Celoria 16, 20133 Milano, Italy}

\begin{abstract}
By means of the exact Path Integral Monte Carlo method we have performed a detailed microscopic study of \4He nanodroplets doped with an argon ion, Ar$^+$, at ${T=0.5}$ K.
We have computed density profiles, energies, dissociation energies and characterized the local order around the ion for nanodroplets with a number of \4He atoms ranging from 10 to 64 and also 128.
We have found the formation of a stable solid structure around the ion, a ``snowball'', consisting of 3 concentric shells in which the \4He atoms are placed on at the vertices of platonic solids:
the first inner shell is an icosahedron (12 atoms); 
the second one is a dodecahedron with 20 atoms placed on the faces of the icosahedron of the first shell; the third shell is again an icosahedron composed of 12 atoms placed on the faces of the dodecahedron of the second shell.
The ``magic numbers'' implied by this structure, 12, 32 and 44 helium atoms, have been observed in a recent experimental study~\cite{Bartl14} of these complexes;
the dissociation energy curve computed in the present work shows jumps in correspondence with those found in the nanodroplets abundance distribution measured in that experiment, strengthening the agreement between theory and experiment.
The same structures were predicted in Ref.~\cite{Galli11} in a study regarding Na\+@\4He\n when $n>30$; a comparison between Ar\+@\4He\n and Na\+@\4He\n complexes is also presented.


\pacs{67.25.dw}

\end{abstract}

\maketitle

\section{Introduction}
\label{intro}

The use of charged impurities in bulk \4He has a long history as they constitute one of the first microscopic probes employed in the study of superfluid phenomena in liquid \4He~\cite{Fetter76}.
In particular, positive ions are believed to generate in the solvent a region of increased density as a result of the electric field originating from its charge. 
The local density is estimated to be so large that \4He should solidify around the ion. This is the phenomenological ``snowball'' model developed by Atkins~\cite{Atkins59} and widely used in the interpretation of the earlier experiments of ions in bulk superfluid \4He. 
A similar snowball is expected to be present also if the ion is captured inside a droplet of \4He; in fact, many experiments performed with various positive ions have been interpreted in terms of the formation of a snowball~\cite{Toennies98,Barranco06,Tiggesbaumker07}.
Nanodroplets of helium have been extensively studied in the last two decades due to their fundamental properties and the applications that they provide; helium droplets can act as a matrix where neutral or charged molecules and atoms and chemical reactions~\cite{Lugovoj00,Krasnokutski11} can be investigated by means of high resolution spectroscopy~\cite{Toennies98,Whaley98,Toennies01,Callegari01,Stienkemeier01}. Moreover, they provide unique environments to investigate superfluidity in finite systems via rotation of embedded dopants~\cite{Toennies98,Toennies01,Callegari01,Stienkemeier01,Barranco06}.    

Quantum Monte Carlo (QMC) methods have been extensively used in the past to obtain microscopic studies of ions in bulk and in nanodroplets of \4He~\cite{Buzzacchi01,Galli01,Rossi04,Barranco06,Coccia07,Paolini07,Coccia08,Galli11}.
These studies confirmed the presence of a snowball around the ion characterized by the formation of well-defined shells of \4He atoms; the size of the snowball, the number of atoms in the shells and the kind of local order were found to depend on the specific ion. This behavior is different from the one predicted by the Atkins model, where the snowball depends only on the ionic charge.

A recent experimental study, of Bartl et al.~\cite{Bartl14}, investigated \4He nanodroplets doped with noble gases ions, Ar\+, Kr\+ and Xe\+, and observed a rich spectrum of anomalies in the abundance distribution indicating the formation of different types of structures.
In particular, the Ar\+@\4He\n complexes showed evidence for three distinct solvation shells containing, respectively, 12, 20 and 12 helium atoms, going from the first to the third shell. We note, moreover, that the evidence of a first shell of 12 more strongly bonded helium atoms surrounding Ar\+ was also found in earlier experimental works~\cite{Kojima92,Callicoatt98,Brindle05,Kim06,FerreiradaSilva09}. The same three magic numbers have been observed in a path integral Monte Carlo (PIMC) study of Na\+@\4He\n complexes with a number of atoms greater than 30~\cite{Galli11}; such numbers correspond to a highly ordered snowball with three ordered shells compatible with an icosahedron, in the first and third shells and a dodecahedron in the second shell. The purpose of the present work is to verify with a microscopic study the experimental evidences observed in Ref.~\cite{Bartl14}.
To pursue our aim, we have performed PIMC simulations of the Ar\+@\4He\n complexes at ${T=0.5~\mathrm{K}}$ for numbers, $n$, of \4He atoms ranging from 10 to 64 and also 128; we  derived density profiles, total and dissociation energies as function of the number of \4He atoms. We have also performed a microscopic analysis of the local order in the solvation shells. 

The article is organized as follows: in Sec. 2 we give the Hamiltonian of the system and some details of our PIMC simulations; in Sec. 3 we present and discuss our results; in the last Section we draw our conclusions and perspectives.

\begin{figure}[t]
\begin{center}
\includegraphics[width=0.495\textwidth]{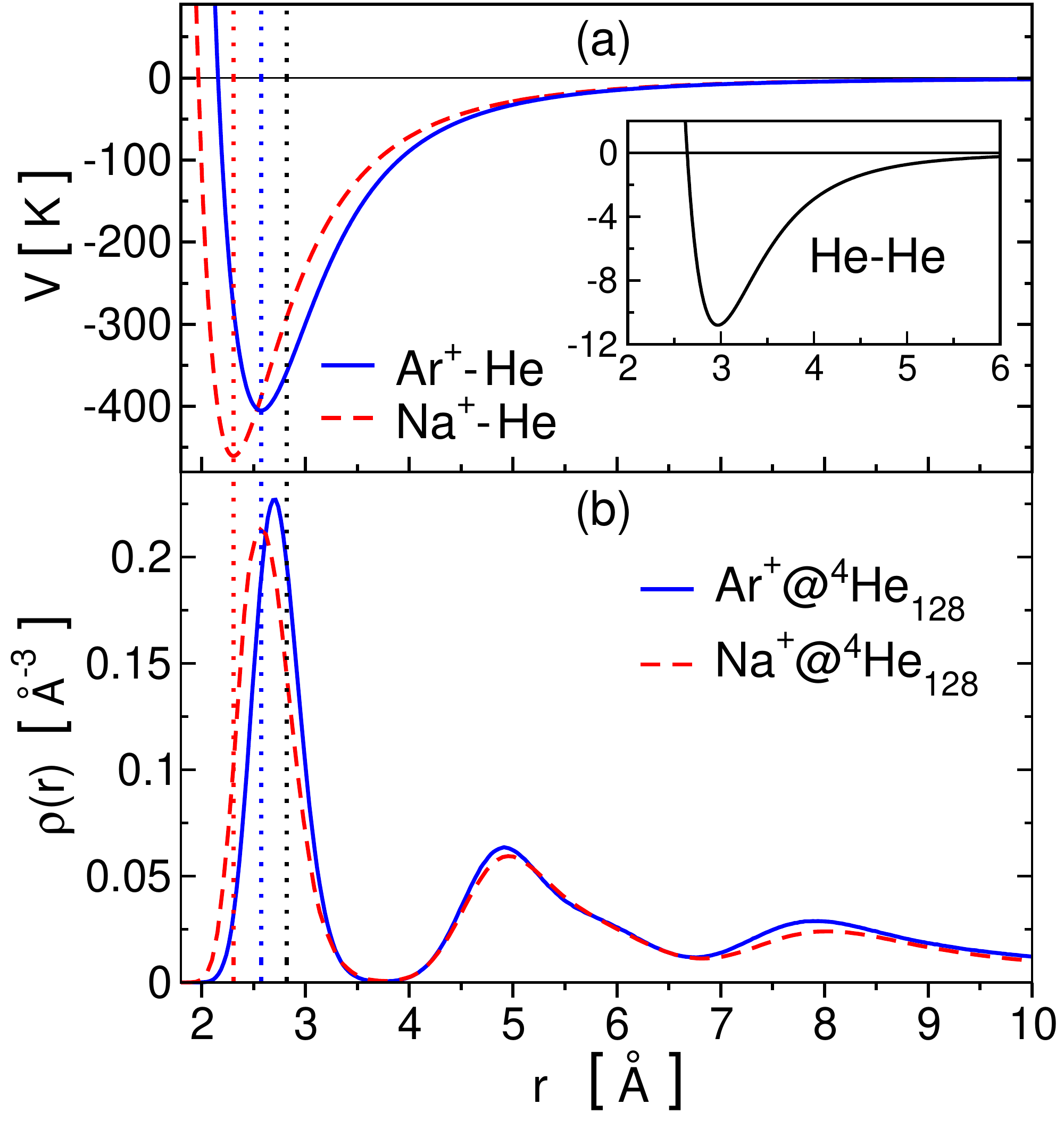}
\includegraphics[width=0.495\textwidth]{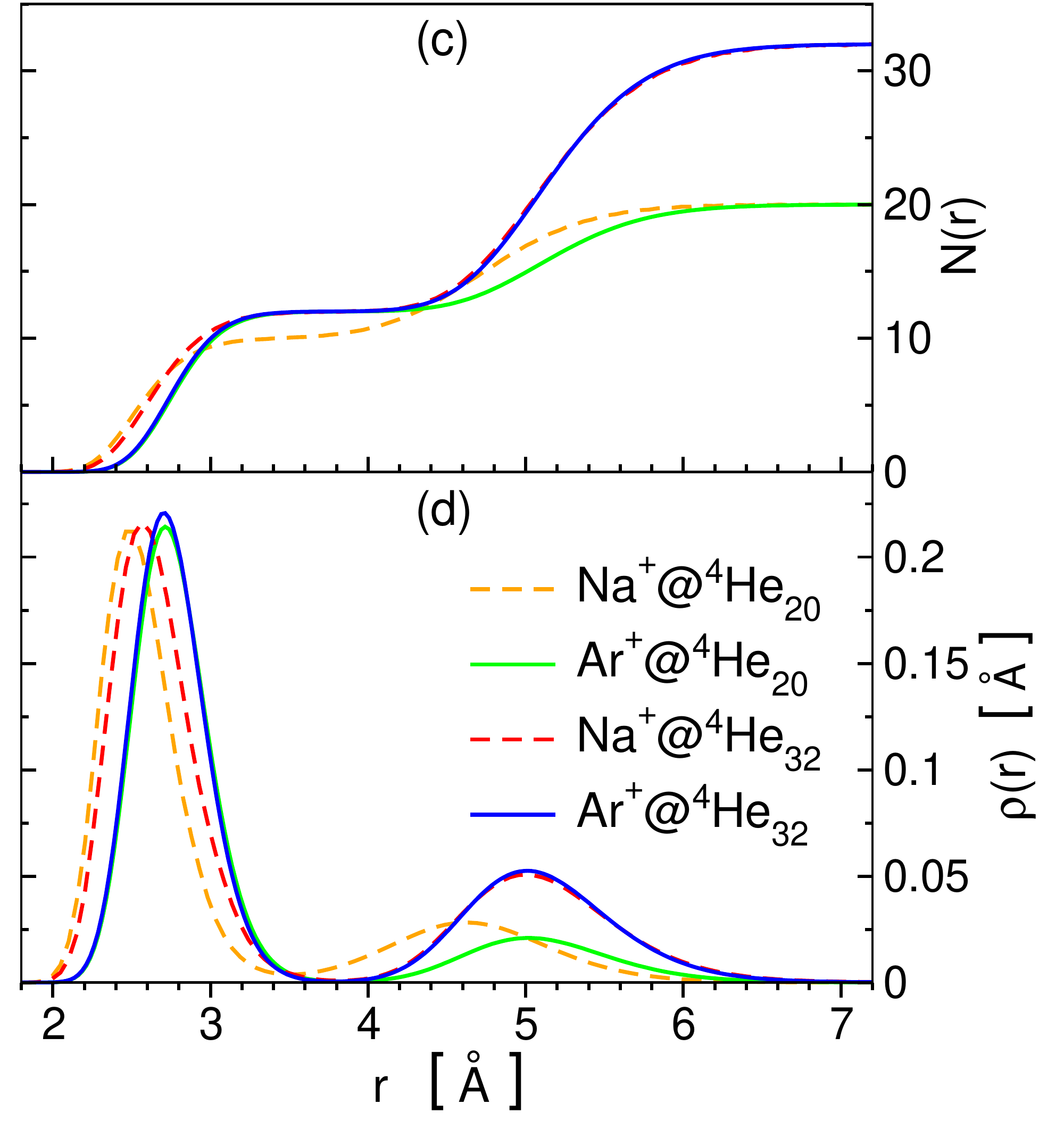}
\caption{
(a) Ar\+--He~\cite{Carrington95} (solid line) and Na\+--He~\cite{Ahlrichs88} (dashed line) interaction potentials compared with the He--He interaction potential~\cite{Aziz79} shown in the inset.
The dotted vertical lines show the positions of the minimum of the ion--He potentials (blue and red) and of the ideal ``distance'', $r_{id}$, (black) (see text). (b) Radial \4He density, $\rho(r)$, around Ar\+ and Na\+ ions in nanodroplets with 64 and 128 \4He atoms.
(c) Number of \4He atoms, $N(r)$, within a sphere of radius $r$ around the ion for nanodroplets Ar\+@\4He\n with $n=$20, 32.
(d) Radial \4He density, $\rho(r)$, around Ar\+ and Na\+ ions in nanodroplets with 20 and 32 \4He atoms. (color figure online)
 }
\label{fig:pot+dens}
\end{center}
\end{figure}

\section{Model and methods}
In this study we have employed the PIMC method ~\cite{Pollock87,Ceperley95} using a canonical variant of the
Worm algorithm~\cite{Boninsegni06a,Boninsegni06b,Mezzacapo06,Mezzacapo07}.
As short--time density matrix we have utilized a pair approximation for the multi--product expansion density matrix (pair--MPE): this corresponds to a pair action obtained by using the multi--product expansion (MPE) approximation~\cite{Zillich10} for the two body propagator and by approximating the many--body density matrix as a product of two--body terms.
As time step, we used $\delta\tau=\SI{1/160}{K^{-1}}$ .

The Hamiltonian that describes the system has the form
\begin{equation}
\label{eq:hamiltonian}
\hat{H}=
-\frac{\hbar^{2}}{2m_{4}} \sum_{i=1}^{n} \nabla^{2}_{i} 
-\frac{\hbar^{2}}{2m_{Ar}} \nabla^{2}_{Ar} + 
\sum_{i=1}^{n} v_{Ar}(|\vec{r}_{Ar}-\vec{r}_{i}|)  + \!\!\!
\sum_{1\leq i<j\leq n} \!\!\! v_{4}(|\vec{r}_{i}-\vec{r}_{j}|)
\end{equation}
where $m_4$ is the mass of a \4He atom, $m_{Ar}$ is the mass of the ion Ar\+, $v_4$ is the He--He interaction potential~\cite{Aziz79}, $v_{Ar}$ is the the Ar\+--He interaction potential~\cite{Carrington95}, $\{\vec{r}_i \}_{i=1}^n$ and $\vec{r}_{Ar}$ are the coordinates of the \4He atoms and of the Ar\+ ion.

In Fig.~\ref{fig:pot+dens}a, we plot the adopted Ar\+--He potential and, by comparison, also the Na\+--He interaction potential used in Ref.~\cite{Galli11}.
It can be noticed that the two ion-He potentials have wells with similar depth and minimum position; the wells are more than 40 times deeper and placed at shorter distances than that of the He--He potential, which is shown in the inset.

The positions of the \4He atoms in the snowball are a result of the competition of the ion-He and He-He interactions against the zero point and thermal motion of the \4He atoms and the ion.
A specific solid structure in the first shell around the ion, with a given number of \4He atoms, corresponds to a minimum of the potential energy when the distances among \4He atoms and the distances among the ion and the \4He atoms correspond to the minimum of the respective interaction potentials.
In a perfect icosahedron, with one \4He atom located at each vertex and in which the distance between adjacent \4He atoms is equal to the distance of the minimum of $v_4$ (${r_{\mathrm{He}} \simeq \SI{2.97}{\angs}}$), the vertices would be placed at the ``ideal'' distance ${r_{id} \simeq \SI{2.82}{\angs}}$ from the icosahedron center. 
As shown in Fig.~\ref{fig:pot+dens}a, we note that the position of the minimum of $v_{Ar}$ (${r_{\mathrm{Ar}} \simeq \SI{2.57}{\angs}}$) is closer to $r_{id}$ than the minimum in the Na\+--He potential (${r_{\mathrm{Na}} \simeq \SI{2.31}{\angs}}$). 
Moreover, even if the Ar\+--He potential well is about 12\% less deep, its width at half depth is about 11\% wider, and this should reduce the zero--point motion of the \4He atoms in the first shell.
Thus, the overall effect should be the reinforcement of the stability of the icosahedral geometry in the first shell.
During the simulations, we have verified that $v_{Ar}$ is attractive enough to bind the \4He atoms to the nanodroplets when $T=0.5$ K, therefore our results are not affected by any evaporation of \4He atoms.

\section{Results}
In Fig.~\ref{fig:pot+dens}b we show the radial density profiles, $\rho(r)$, of the Ar\+@\4He$_{n=128}$ complex and, by comparison, also that of the Na\+@\4He$_{n=128}$ complex obtained in Ref.~\cite{Galli11}. 
The density profiles for Ar\+ and for Na\+, apart from the position of the first radial peak, $r^{peak}$, are very similar.
In the Ar\+ case, the first peak is placed farther away from the ion, as an effect of the different position of the minimum of the potential well.
However, the ratio ${r_\mathrm{Ar}^{peak}/r_{\mathrm{Ar}}\simeq 1.05}$ is closer to one than the ratio $r_\mathrm{Na}^{peak}/r_{\mathrm{Na}}\simeq 1.12$, indicating that the first shell in the Ar\+@\4He$_{n=128}$ nanodroplet follows more closely the minimum of the He--ion potential.
We note also that the shape of the second modulation in the density profile is almost identical to that obtained in the Na\+ case. 
As in the Na\+ case~\cite{Galli11}, this modulation turns out to correspond to two tightly neighboring shells, the second and the third.
\begin{figure}[t]
\begin{center}
\includegraphics[width=0.75\textwidth]{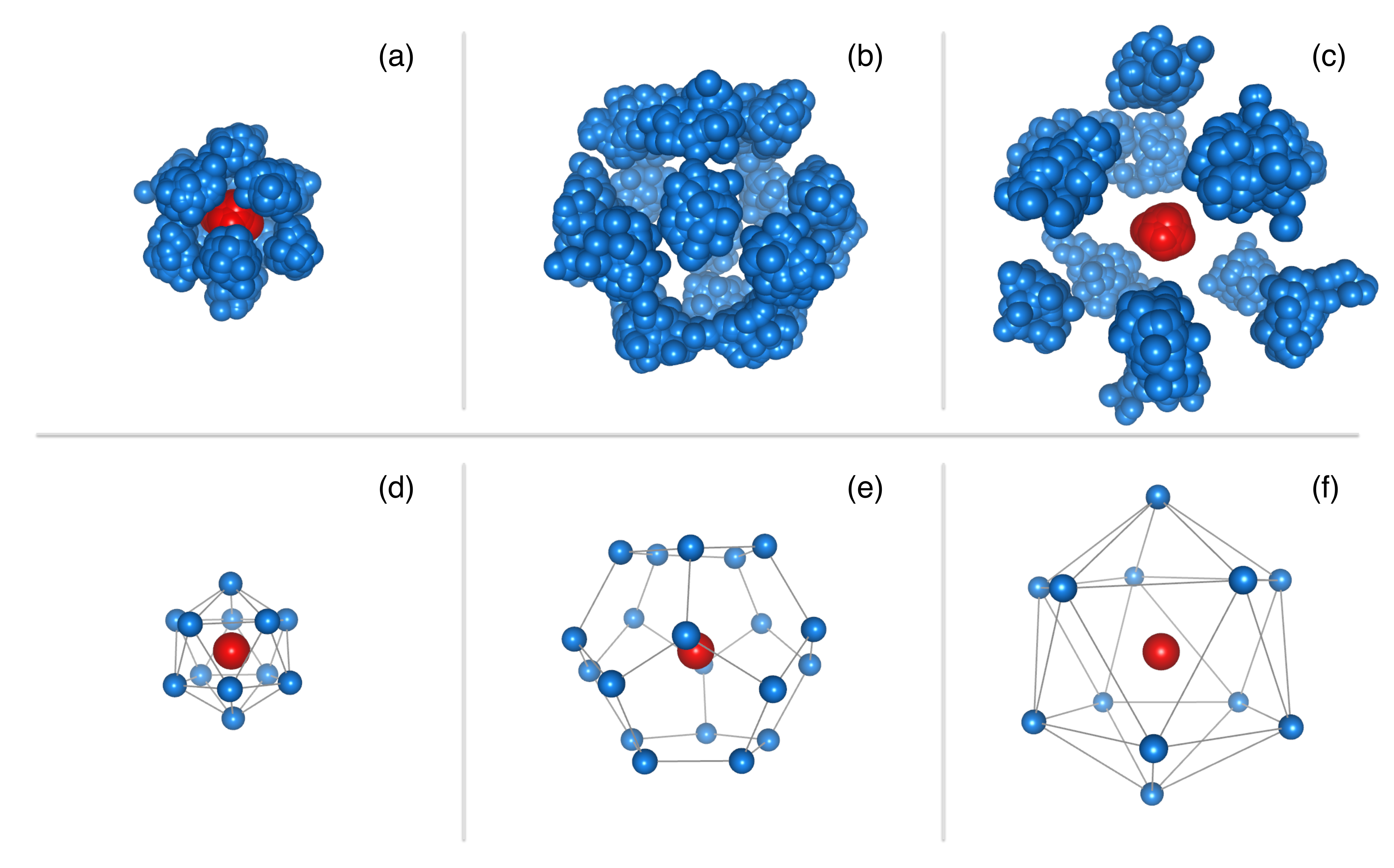}
\caption{Example of a spatial configuration of polymers for a simulated nanodroplet with 128 \4He atoms.  
In each of the top panels are shown the polymers associated to \4He atoms of the first (a), the second (b) or the third shell (c) and the polymer of the Ar\+ ion (red balls).
In the bottom panels (d,e,f) are shown the same shell configurations of the top panels, but each polymer is replaced its center of mass. (color figure online)}
\label{fig:shells}
\end{center}
\end{figure}

To determine which type of local solid order characterizes the shells, we have  investigated the spatial configurations of polymers produced during the simulations.
From each configuration we have selected, by radial position, the polymers placed in one of the three shells.
In the top panels of Fig.~\ref{fig:shells} we show the polymers associated with the Ar\+ ion and the \4He atoms belonging to the first (Fig.~\ref{fig:shells}a), the second (Fig.~\ref{fig:shells}b) and the third (Fig.~\ref{fig:shells}c) shell for a selected configuration. 
In order to better identify the geometric shape of the shells we have also shown the centers of mass of the polymers of the same configuration in the bottom panels of Fig.~\ref{fig:shells}.
One can easily recognize an icosahedron in the first shell and a dodecahedron in the second shell; in the second shell, each atom turns out to be placed on the face of the icosahedron underneath. The third shell is again an icosahedron, where each atom turns out to be placed on the face of the dodecahedron beneath.
From our PIMC simulations we have obtained evidence for a strong stability of the icosahedral geometry in the first shell of Ar\+@\4He$_n$ at low temperatures: we have found that the number of \4He atoms in the first shell is 12 for any studied $n \geq 12$. 
This is different from what was found for Na\+ doped nanodroplets, where only 10 \4He atoms are present in the first shell for $n$ below 30, while 12 atoms occur only for $n$ above 30~\cite{Galli11}. 
This is visible in the panel (c) of Fig.~\ref{fig:pot+dens}, where we show the number of \4He atoms, $N(r)$, contained in a sphere of radius $r$ centered on the ion, obtained through a numerical integration of the radial density profiles (see Fig.~\ref{fig:pot+dens}d) for $n=20, 32$.

\begin{figure}[t]
\begin{center}
\includegraphics[width=0.495\textwidth]{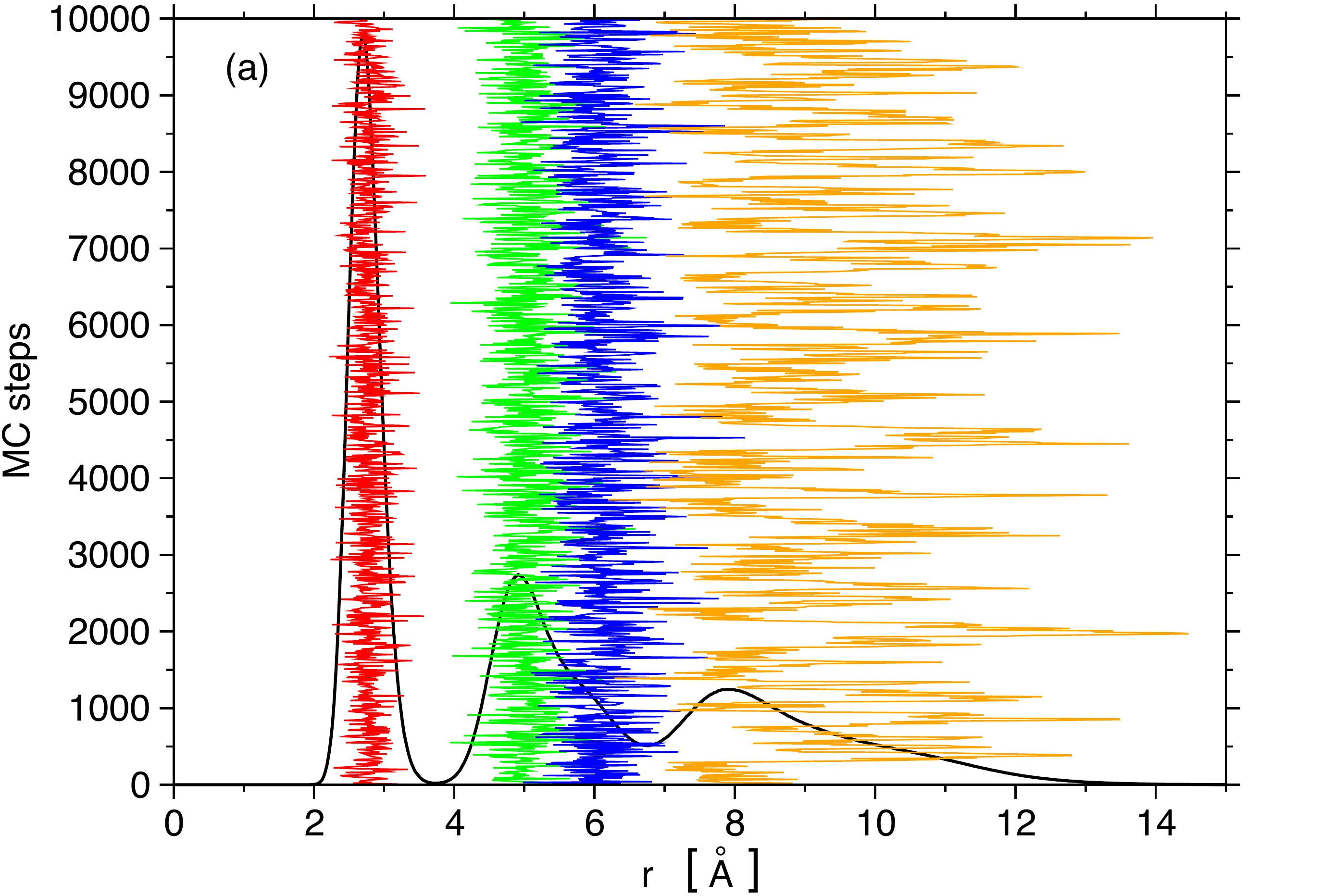}
\includegraphics[width=0.495\textwidth]{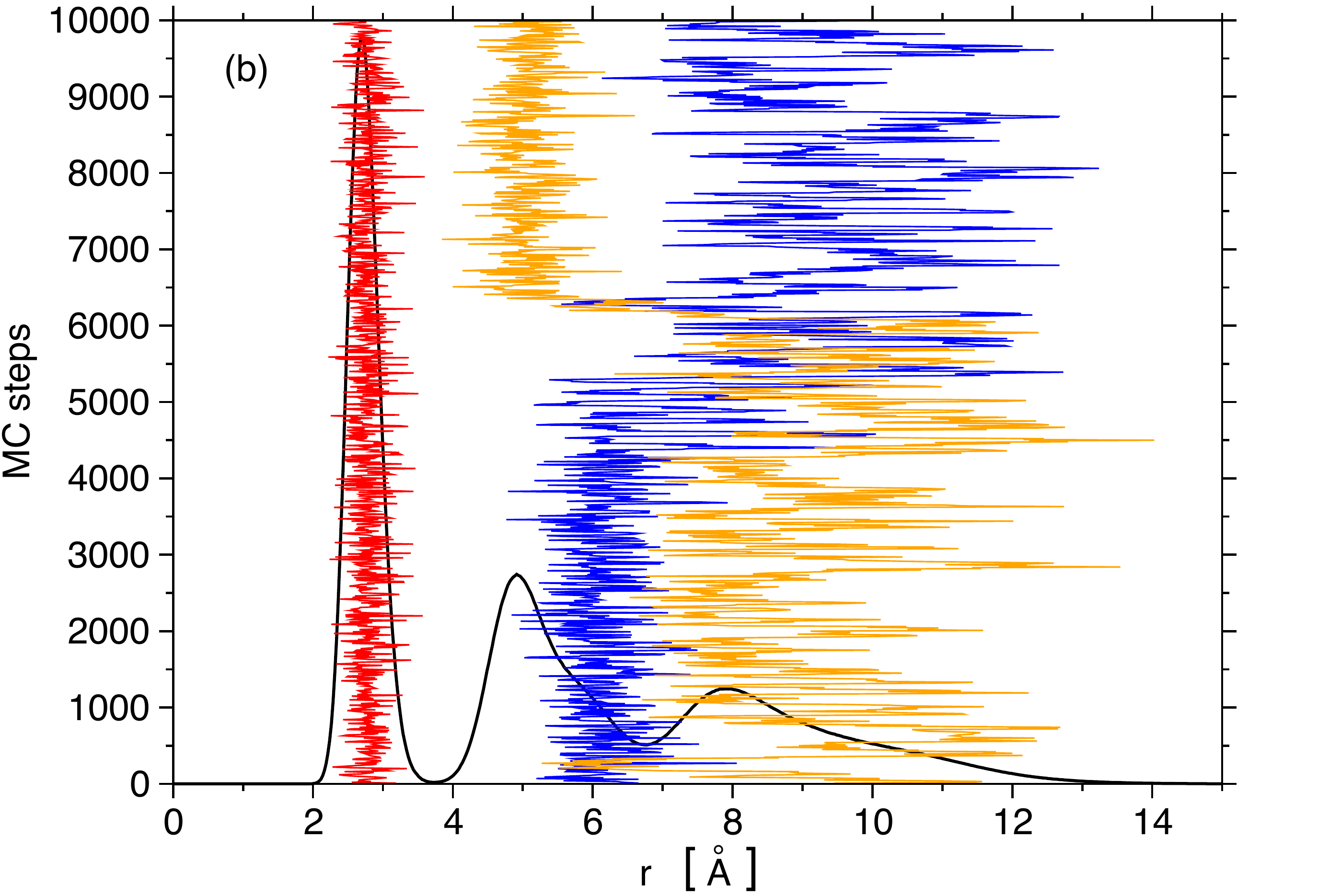}
\caption{Evolution of the distance of some \4He atoms from the Ar\+ ion during the MC simulation. Panels {\bf a} and {\bf b} show examples of such evolutions without (a) and in presence of (b) exchanges between the solid shells and the outer fluid part. (color figure online)}
\label{fig:evolutions}
\end{center}
\end{figure}

To further characterize the microscopic nature of the solvation shells we have analysed also the radial localization of \4He atoms around the Ar\+ ion by measuring the distance between the \4He atoms and the ion during a PIMC simulation.
In Fig.~\ref{fig:evolutions} we show the Monte Carlo evolution of the radial positions of four atoms started at different distances from the Ar\+ ion in a nanodroplet of 128 \4He atoms as function of the Monte Carlo steps; a density profile is also shown in order to easily recognize the presence of jumps among the shells.
During our simulations we observed that atoms started in the first shell (around $r=\SI{2.7}{\angs}$) fluctuate around their starting values, without moving to an outer shell.
Also most of the atoms started in the second (${4.6 \lesssim r \lesssim 5.5~\Ang} $) or in the third shell (${5.5 \lesssim r \lesssim  6.3~\Ang}$) fluctuate around their starting values, but sometimes (see Fig.~\ref{fig:evolutions}b) an atom moves to another shell, or to the external region; conversely, an atom started in the external region can move to the second or third shell.
This analysis confirms the presence of local solid order in the two first modulations of the density profile, modulations which correspond to the first three shells around the Ar\+ ion.
A very interesting issue for the future would be the measurement of the local superfluid density around the ion to investigate the simultaneous presence of solid order and superfluidity; we left such calculations for future investigations.

\begin{figure}[t]
\begin{center}
\includegraphics[width=0.495\textwidth]{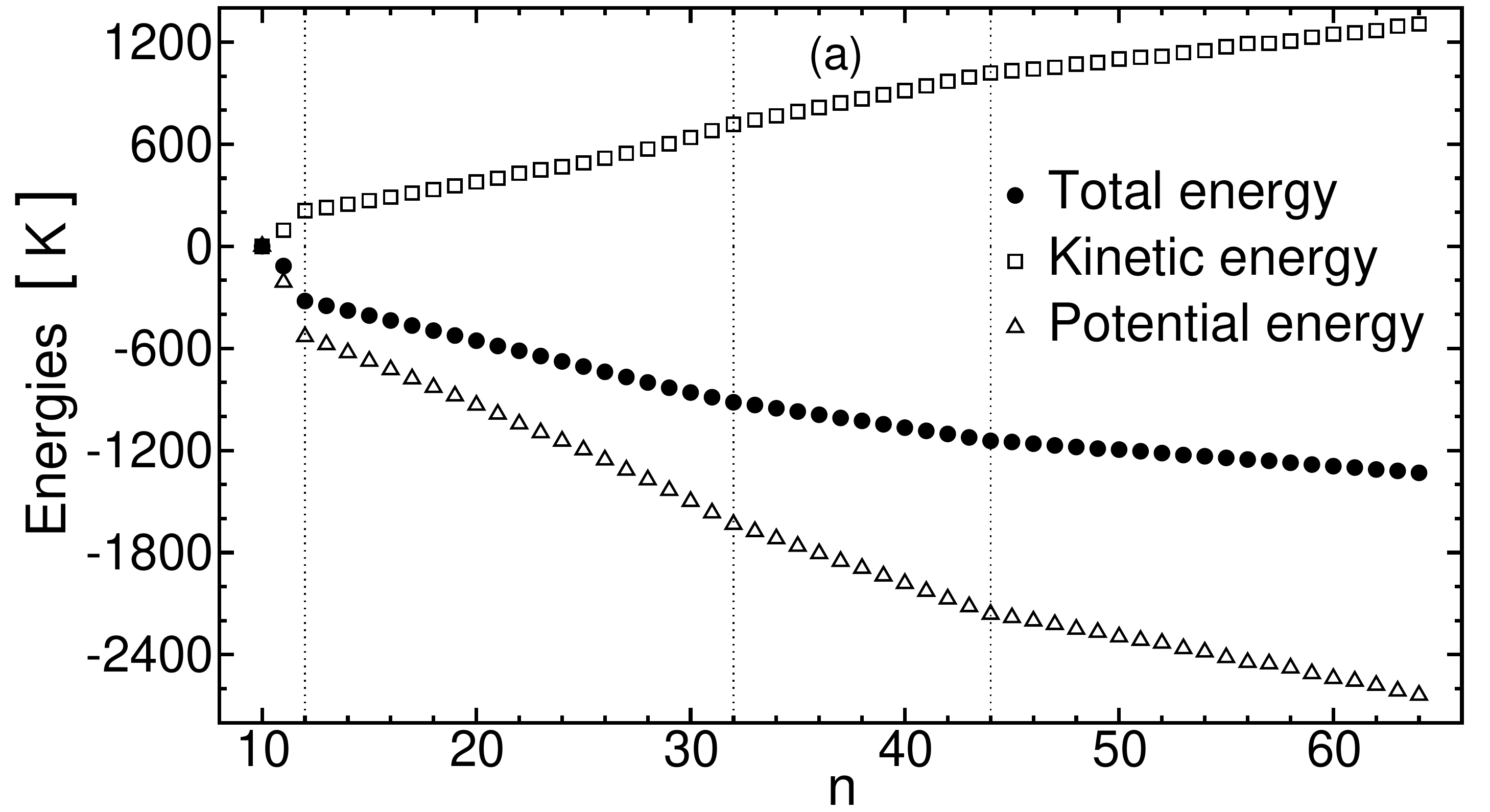}
\includegraphics[width=0.495\textwidth]{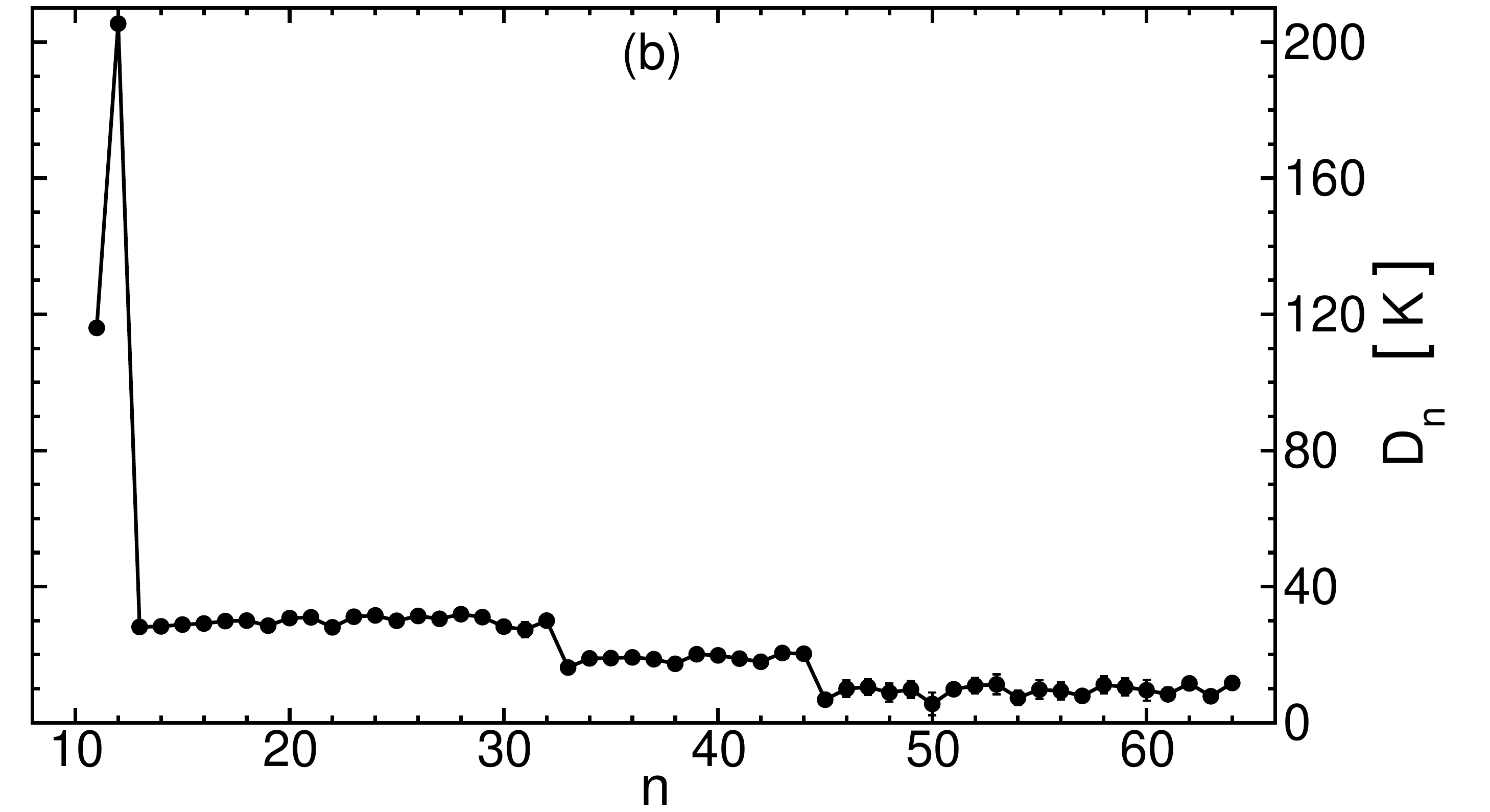}
\caption{
(a) Total $E_n-E_{10}$, kinetic $K_n-K_{10}$ and potential $V_n-V_{10}$ energies per particle of Ar\+@\4He$_n$ complexes with different number, $n$, of \4He atoms at ${T=\SI{0.5}{K}}$. All these energies has been shifted by the respective energies obtained for $n=10$: ${E_{10} =\SI{-2976(1)}{K}}$, ${K_{10} =\SI{614(1)}{K}}$ and ${V_{10}=\SI{-3590.8(1)}{K}}$.
(b) Dissociation energy ${D_n = E_{n-1} - E_n}$.
}
\label{fig:energy}
\end{center}
\end{figure}

One of the experimental approaches to investigate the structure and the size of the snowball is to obtain information on the dissociation energies $D_{n}$, defined as $D_{n}=E_{n-1}-E_{n}$. Rapid drops in $D_{n}$ indicates the closure of solvation shells. 
In order to obtain information on $D_{n}$, one can look for anomalies in the abundance distributions, $I_{n}$, of \4He\n X\+ complexes by mass spectrometry~\cite{Bartl14}. In these experiments the observed complexes are products of dissociations caused by ionization of larger precursor ions. 
The dissociation is a statistical process and the final statistical distribution of the products versus their size, for small droplets, is related to the dissociation energies, $D_{n}$~\cite{Lan12}. 
In this way the anomalies in $D_{n}$, signs of closures of the solvation shells, are reflected in the measured abundance distribution $I_{n}$.
We have calculated the total, kinetic and potential energies $E_{n}$, $K_{n}$, $V_{n}$, of the Ar\+@\4He\n nanodroplets for different numbers of \4He atoms, $n$ (see Fig.~\ref{fig:energy}a). 
One can observe clearly three changes of slope exactly at the magic numbers found in Ref.~\cite{Bartl14}, i.e. for $n=$ 12, 32 and 44.
From these energies we have also computed the dissociation energy $D_{n}$ shown in Fig.\ref{fig:energy}b. Here the magic numbers appear as discontinuity points.
This result confirms the formation of shells with the same number of \4He atoms discussed above, a structure which yields a very stable configuration for the complex.

\section{Conclusions}
In summary, we have performed PIMC simulations of Ar\+ ion doped \4He nanodroplets at low temperatures and characterized the microscopic structure of the snowball. 
We find that the structure around the ion Ar\+ is a highly ordered solid, a \textit{snowball}, composed of three shells in which \4He atoms are placed on the vertices of platonic solids: an icosahedron for the first and the third shells and a dodecahedron for the second shell.
This solid structure was proposed in the interpretation of recent mass spectrometry measurements on Ar\+@\4He\n complexes~\cite{Bartl14}, where three magic numbers, 12, 32, 44, were found in the anomalies in the abundance distribution.
The calculation of the total and the dissociation energy as a function of the size of the nanodroplets confirms the interpretation of the experimental data.

The same structure has been previously predicted in PIMC calculations for \4He\n nanodroplets doped with Na\+~\cite{Galli11} when $n>30$.  
In the Ar\+@\4He\n complexes the icosahedral order in the first shell has been found to be more stable and present for any $n \geq 12$ simulated in the present study.
Remarkably, the analysis of the evolution of the radial positions of the \4He atoms during the PIMC simulations revealed the occurrence of exchanges of helium atoms of the second and third shell with the outer region of the system.
This raises the question of whether or not these atoms participate in local superfluidity.
It will be interesting in the future to calculate local superfluid densities around the ions, especially in the shells where solid order is present.

\begin{acknowledgements}
We acknowledge the CINECA and the Regione Lombardia award, under the LISA initiative, for the availability of high performance computing resources and support.
We acknowledge Prof. Olof Echt for bringing to our attention the results of Ref.~\cite{Bartl14}. 

\end{acknowledgements}

\bibliography{Ref_Clusters}   


\end{document}